\documentstyle[prd,aps]{revtex}
\begin{document}
\draft
\input epsf
\twocolumn[\hsize\textwidth\columnwidth\hsize\csname
@twocolumnfalse\endcsname

\title{$D$-branes, String Cosmology and Large Extra Dimensions}
\author{Antonio
Riotto}

\address{{\it Theory Division, CERN, CH-1211,
Geneva 23, Switzerland}}
\date{April, 1999}
\maketitle
\begin{abstract}

$D$-branes are fundamental in all scenarios where  there are large extra dimensions and the string scale is much smaller than the four-dimensional  Planck mass. We show that this current picture leads to a new approach to string cosmology where inflation on our brane is driven by the large extra dimensions and the  issue of  the graceful exit  becomes 
inextricably linked to the problem of the stabilization of the extra dimensions, suggesting the possibility of a common solution.
We also show  that   branes may violently fluctuate      along their transverse directions in curved spacetime, possibly leading to a period of brane-driven inflation. This phenomenon plays also a  crucial role in many other cosmological issues, such as the smoothing out of the  cosmological singularities and the generation of the baryon asymmetry on our three brane.

\end{abstract}
\vskip2pc]

\def\simlt{\stackrel{<}{{}_\sim}}
\def\simgt{\stackrel{>}{{}_\sim}}

$D$-branes are classical solutions  which are known to exist
in many string theories \cite{pol}. A $Dp$-brane denotes a configuration which extends along $p$ spatial directions and is localized in all other spatial transverse directions. From the string point of view,  a $D$-brane is a soliton
on which open string endpoints may live and 
whose mass (or tension) is proportional to the inverse of the string coupling $g_s$. This  means that $D$-branes become light in those regions of the 
moduli space where the string coupling is large.  This  simple property has led to a new interpretation for singularities of various fixed spacetime backgrounds in which strings propagate and  provided a generic and physically intuitive mechanism for smoothing away cosmological singularities in the strong coupling/large curvature regime of the  evolution of the early Universe \cite{sing1,sing2}. 

The presence of branes in string theory has also given rise to the possibility that the  Standard Model (SM) gauge fields live on  branes rather than in the bulk of spacetime. This scenario corresponds to a novel compactification in which gravity lives in the bulk of the ten dimensional spacetime, but the other observed fields are confined to a brane of lower dimension \cite{witten}. 
In particular, one might imagine the situation where the SM particles live on a three brane and the directions transverse to the three brane being compact. There might be also other branes separated from ours
in the extra directions. Since the usual gauge and matter fields live on the brane, the only long-range interaction which sees the extra dimensions is gravity. 

In the  weakly coupled heterotic string there is a fundamental relation between the 
 string scale $M_s$, the four-dimensional Planck mass $M_p$, the value
of the dilaton field $\phi$ and the gauge coupling constant $\alpha_g$, $(M_s/M_p)^2\sim\alpha_g\sim e^{\phi}$ and the string scale
is
fixed to be about $10^{18}$ GeV. However, in other regions of the
moduli space the string scale can be much smaller \cite{witten} and it is tempting to imagine \cite{lykken} that the string scale is not far from the
TeV region. If this is the case, the small ratio $M_s/M_p$ could be a consequence of  large
compactified dimensions \cite{large1,large2} through a scaling relation like

\begin{equation}
M_p^2 = M_{s}^{n + 2}V_n, \label{r}
\end{equation}
where $V_n$ is the volume of the $n$ large extra dimensions. For $M_s\sim$ TeV and $n\geq 2$, the radius of the large extra dimensions may be as large as the 
millimeter. Large extra dimensions are only allowed if
the SM fields are confined to live on a membrane orthogonal to the
large extra dimensions and this hypothesis fits rather nicely with the
current picture of string theory and $D$-branes.

Not surprisingly, one expects many interesting cosmological implications of  this scenario  (some of them have already been discussed in \cite{sing1,cosmo}) and the possibility of large extra dimensions is potentially exciting for several reasons. 

In the presence of some matter
with energy density $\rho$, the evolution of the Universe is controlled by the equation $H^2\equiv (\dot{a}/a)^2=(8\pi\rho/3 M_p^2)$, where $a$ is the scale factor and $H$ is the Hubble rate. This implies that for an expanding Universe, $\rho$ and $H^2$ are proportional, they increase or decrease together in time.
This is true only if $M_p$ is constant.  Now, in the weakly coupled string theory where all compactified dimensions have size $\sim M_s^{-1}$, $M_p$ is controlled by the dilaton and it is perfectly possible to have growing $H$ while $\rho$ decreases, provided the dilaton $\phi$ is also 
growing. On this peculiar feature is based the whole idea of Pre Big-Bang (PBB) cosmology where there exist superinflationary solutions for the scale factor $a$ having  a growing -- rather than a constant -- Hubble parameter and a growing dilaton field \cite{ven}. The suggestion from weakly coupled string theory is therefore rather compelling, since it is possible to generate a  scenario containing a dilaton-driven PPB inflation. 

These  simple considerations together with the relation (\ref{r}) holding  in the current picture of string theory  suggest  a simple way of getting inflation on our three brane and a fresh approach to string cosmology: since $M_p$ is now governed by the size $b$ of the extra dimensions, it is possible
to obtain superinflationary solutions provided that the extra dimensions change in time, that is a radion-driven PPB cosmology. The Einstein action in
$4+n$ dimensions and metric $g_{MN}=(1,-a^2\delta_{ij},-b^2\delta_{mn})$ reads
\begin{equation}
\label{SE}
S\sim
\int d^4x\, a^3b^n\left[ R-2n\frac{\ddot{b}}{b}-6n\frac{\dot{a}}{a}
\frac{\dot{b}}{b}-n(n-1)\frac{\dot{b}^2}{b^2}\right],
\end{equation}
where $R$ is the usual four-dimensional Ricci scalar. Setting $b^n=e^{c\sigma}$, with  $c=\left(n/n-1\right)^{1/2}$, and integrating by parts the term $\ddot{b}$ in (\ref{SE}), the action for the radion field $\sigma$ becomes
\begin{equation}
S\sim\int d^4x a^3\, e^{c\sigma}\left[ R +
\partial_{\mu}\sigma\partial^{\mu}\sigma\right].
\end{equation}
Were  $c=1$, string cosmology experts 
would recognize the action of the dilaton field coupled to gravity
(with $\phi=-\sigma$). For us  $c=1$ in the limit of   $n\rightarrow\infty$ and in this regime we can recover the more familiar results of dilaton-driven PPB cosmology. The evolution of the  radion is governed by the equation obtained by varying the lapse function
\begin{equation}
\dot{\sigma}=\epsilon_{\pm}H,\;\;\epsilon_{\pm}=-3c\pm\sqrt{9c^2-6}, 
\end{equation}
while the equation for the Hubble parameter becomes
\begin{equation}
\dot{H}=-\alpha_{\pm}H^2,~~~\alpha_{\pm}=3+c\epsilon_{\pm}.
\end{equation}
For each value of $n$, $\alpha_{-}$ is negative and it generates a superinflationary solution, that is $H$ grows until a singularity is reached, say at $t=0$.  In terms of the scale factors $a$ and $b$, the solutions
are given by $a(t)=a_0\left(t/t_0\right)^{-\gamma_1}$ and $
b(t)=b_0\left(t/t_0\right)^{\gamma_2}$ where
$t$ and $t_0$ are negative and $\gamma_1=|1/\alpha_-|$,
$\gamma_2=|c\epsilon_-/(n\alpha_-)|$. Since  $\epsilon_-<0$, the volume of the internal dimensions is contracting. Suppose that $a$ and $b$ start with  comparable values at low curvature at $t\sim t_0$, say $a\sim b\sim M_s^{-1}$; the subsequent  evolution is such that the spatial dimensions of our three brane superinflate, whereas
the extra dimensions evolve to values smaller than $M_s^{-1}$. In string theory, due to the presence of winding states in the closed string (gravitational)
sector, the low-energy physics is equivalent as if there was a radius 
$b_I=
(b M_s^2)^{-1}$
which is much larger than the string length and the roles
of winding and Kaluza-Klein momenta are interchanged (the evolution in terms of $b_I$  is also characterized  by a nontrivial dynamics for the dilaton field). On the other hand, in this T-dual description, the open string states --  which give rise to the ordinary 
non gravitational matter on the three brane -- have only heavy winding modes. As a result, in the dual theory the extra dimensions appear to be  large.
The dynamics of the radion field $b$ coupled to the four-dimensional gravity leads automatically to an inflationary stage on our three brane and explains 
the huge hierarchy between the sizes of our observed Universe and  
of the extra dimensions if $|\gamma_1|\gg |\gamma_2|$. The cosmological solution stops being valid when 
the Hubble rate becomes of the order of $M_s$. However, differently from the
dilaton-driven PPB cosmology where one has to face a regime of strong coupling
when approaching the big-bang singularity at $t\rightarrow 0$, in the radion-driven PPB scenario the singularity may take place well inside the pertubative regime and be understood as the process
of decompactifying the extra dimensions. The
 issue of  the graceful exit in string cosmology    becomes therefore 
inextricably linked to the problem of the stabilization of the extra dimensions, suggesting the possibility of a common solution.

 There is -- however -- another feature of $D$-branes that may help in solving the radion-driven PPB  singularity problem as well as give new insight into other fundamental cosmological issues: this  is the peculiar behaviour of fluctuations transverse to the brane in curved spacetime. The action governing the dynamics of a generic 
$Dp$-brane (neglecting the gauge fields) is~\cite{pol} $
S=-T_p\int d^{p+1}\xi\, e^{-\phi}\sqrt{-
{\rm det} G_{\alpha\beta}}$,
 where 
 $T_p^{-1}$ is the brane tension, 
$\xi^{\alpha}$ ($\alpha=0,\ldots ,p$)
parametrize the $D$-brane world-volume and  
$G_{\alpha\beta}=g_{\mu\nu}(X)\partial_{\alpha} X^{\mu}\partial_{\beta} X^{\nu}$ is the induced metric.  Given a brane configuration $\bar{X}^\mu(\xi)$ in a spacetime of $D$-dimensions,  arbitrary 
fluctuations transverse to
this configuration (the would-be Goldstone bosons of the  broken translational invariance along the directions perpendiclar to the brane)  are parametrized by $(D-p)$ world-volume scalars
$y^a(\xi )$ as $
X^{\mu}(\xi)=\bar{X}^{\mu}(\xi )+\bar{n}^{\mu}_a(\xi )y^a(\xi )$,
where $\bar{n}_a^{\mu}(\xi )$ ($a=1,\ldots ,(D-p)$) are the $(D-p)$ 
vectors normal to the brane at the point
$\bar{X}(\xi )$.
The expansion of the metric in Riemann
coordinates is~\cite{Pet} 
\begin{equation}
\label{exp}
g_{\mu\nu}(X)=\bar{g}_{\mu\nu}-\frac{1}{3}y^a y^b
\bar{R}_{\mu\rho\nu\sigma} \bar{n}_a^{\rho} \bar{n}_b^{\sigma}
+{\cal O}(y^3),
\end{equation}
where the overbar indicates that the quantity is evaluated at
$\bar{X}$. If the extrinsic curvature of the membrane is small, the
normal vectors are 
$\bar{n}_a^{\rho}\simeq\delta_a^{\rho}$, and therefore $
g_{\mu\nu}(X)=\bar{g}_{\mu\nu}-\frac{1}{3}y^a y^b
\bar{R}_{\mu a\nu b}+{\cal O}(y^3)$ \cite{sing1}.

Let us begin, for simplicity, by considering a single $Dp$-brane in an expanding accelerating ($\ddot{a}>0$) background
where the dimensions orthogonal to the brane are static. We shall consider
for definiteness a background where the common scale factor $a$ of the spatial
coordinates $\vec{\xi}$  is inflating,  $a(\eta)\sim (-\eta)^{-q}$, where $0<q\leq 1$ and $d\eta=(dt/a)$ is the conformal time, $-\infty<\eta<0$. For 
$q\neq 1$ the Hubble rate $H$ grows with time till a singularity is reached and the De Sitter background with constant $H$ is recovered for $q=1$. The equation of motion  for  small transverse oscillations
$y_a(\xi)$ ($y_a$ is the comoving displacement) is
\begin{equation}
\ddot{y_a}+ (p-1)(\dot{a}/a)\dot{ y_a}-\nabla^2 y_a=0,
\end{equation}
where $\dot{ y_a}\equiv \partial y_a/\partial\eta$. The canonical conjugate momentum is $\pi_a=\tau_p a^{p-1}\dot{y}_a$, where $\tau=T_p/g_s$; imposing the canonical commutation relation $[y_a(\eta,\vec{\xi}),\pi_a(\eta,\vec{\xi}^\prime)]=i\delta^{(p)}(\vec{\xi}-
\vec{\xi}^\prime)$, we can expand the solution as
\begin{equation}
y^a(\eta,\vec{\xi})=\frac{1}{\tau^{\frac{1}{p+1}}}\int\frac{d^pk}{(2\pi)^{p/2}}\left[a_k h^a_k(\eta) e^{-i\vec{k}\cdot\vec{\xi}}+
a_k^\dagger h^{a\dagger}_k(\eta) e^{+i\vec{k}\cdot\vec{\xi}}\right],
\end{equation}
where $a_k$ and $a_k^\dagger$ are the annihilation and creation operators, respectively. The solution is very simple, $h_k^a(\eta)= (-\eta)^\beta\left[c_1 H_\beta^1(-k\eta)+c_2 H_\beta^{1*}(-k\eta)\right]$, where $\beta= (q(p-1)+1)/2$. If we quantize the modes and define the vacuum state by $a_k|0\rangle=0$, then different choices of vacuum correspond to different choices of $c_1$ and $c_2$. The adiabatic vacuum corresponds to $c_2=0$ and this (Heisenberg) state is the state we will assume the brane is in.

The behaviour of a mode $h_k^a(\eta)$ is very simple. Suppose from now on to consider the case $p=3$, {\it i.e.} we consider a three brane. As long as the physical wavelength $ak^{-1}$ of a mode is inside the Hubble radius,  $h_k^a(\eta)$ oscillates with constant
{\it physical} amplitude $ah_k^a(\eta)$. As the physical wavelength grows it crosses the Hubble radius and the {\it comoving} amplitude $h_k^a(\eta)$ becomes frozen, so that the physical amplitude grows like $a$.

In this state one can calculate the mean square transverse displacement
\begin{eqnarray}
\label{mean}
\langle {\bf y}^2\rangle&\equiv & \int \frac{d^3\xi}{V_3}\langle 0|{\bf y}^2(\eta,
\vec{\xi})|0\rangle\nonumber\\
&=&\frac{D-4}{\tau^{1/2}}\int \frac{d^3 k}{(2\pi)^3} \left|h_k^a\right|^2,
\end{eqnarray}
where ${\bf y}^2\equiv \sum_a y_a^2$ and $V_3=\int d^3\xi$.  
The physical displacement
$a{\bf y}$ has the usual  ultraviolet divergence as in flat space: 
 physical wavelengths with     $k\gg a H$  are always well inside the horizon  at a given time and their amplitude is unaffected by the expansion. Their contribution leads to the usual
$\int d^3k k^{-1}$  divergence. We can subtract this divergence.  
Modes with $k\ll a_i H_i$, where $a_i$ and $H_i$ are the  values of the scalar factor and Hubble rate at the beginning of the inflationary stage, are always well outside the horizon and simply match on adiabatically to the modes before and after inflation. One therefore  finds  at $\eta\simeq \eta_f\simeq 0$ (for $q\neq -1$) 
\begin{eqnarray}
\label{just}
a^2\langle {\bf y}^2\rangle &\simeq & a^2 \frac{D-4}{\tau^{1/2}}\int^{a_f H_f}_{a_i H_i} H \frac{d^3 k}{(2\pi)^3} \left|h_k^a\right|^2
\nonumber\\
&\sim & \frac{D-4}{M_s^{4}\eta_f^2},
\end{eqnarray}
where we have set $\tau\sim M_s^4$.  One can picture this result by saying that the modes with wavelength of the order of the Hubble radius have a physical width $\langle {\bf y}_p^2\rangle\simeq M_s^{-2}$ which gets amplified by the expansion after they cross the Hubble radius. Transverse fluctuations grow in time.
In the special case of De Sitter with constant rate of expansion $H$, one finds  that $\langle {\bf y}^2\rangle\simeq M_s^{-4} H^{2}{\rm ln}(EH)\sim M_s^{-4}H^{3}t$, where $E$ is the total number of {\it e}-foldings.

More interestingly, one can calculate the energy acquired by each mode in this process. The energy is given by $
E\sim M_s^4\int d^3\xi a^3\left[1+ 1/2\dot{{\bf y}}^2/a^2+
1/2\left(\nabla{\bf y}\right)^2/a^2\right]$, 
where the first term is just the classical stretching. Just as in (\ref{just}),
we find 
\begin{equation}
\frac{\langle\left(\nabla{\bf y}\right)^2\rangle}{a^2}\simeq \left(\frac{H}{M_s}\right)^4.
\end{equation}
The term $\langle \dot{{\bf y}}^2\rangle$ gives no contribution after the flat spacetime substraction. Thus we deduce that the fractional energy in the transverse perturbations grows with time as $H$  increases with time and their
contribution to the energy density becomes sizeable when $H\sim M_s$. Physical transverse
fluctuations
of the three brane -- even though initially oscillating -- will eventually grow  
and give rise to an instability. Therefore,   as the three brane is stretched by the
expansion of the
Universe, transverse fluctuations grow, the thickness of the brane gets larger and larger  and the brane cannot be
considered as 
a static object  any longer. This growth is accompanied by a huge
production of light  pseudo-Goldstone bosons of the broken translation
invariance. Let us try to feed back the effect of this production into the expansion of the Universe.  As the system evolves towards the singularity, more and more energy in transferred into transverse fluctuations of the brane till -- eventually -- they dominate the energy density of the Universe.  Since $\langle \left(\nabla{\bf y}\right)^2\rangle\gg \langle \dot{{\bf y}}^2\rangle$, the gas of pseudo-Goldstone bosons is characterized by an energy density $\rho_y\simeq
-3 p_y$, where $p_y$ is the pressure of the gas. We thus obtain an effective negative pressure. Assuming a flat Universe, the equation for the acceleration $\ddot{a}$ alone becomes 
\begin{equation}
\ddot{a}/a\propto(\rho_y+3 p_y)=0
\end{equation}
and we discover that the back-reaction  of the gas of pseudo-Goldstone bosons onto the evolution of the Universe is to halt
the period of  acceleration and -- therefore -- to smooth the singularity away when $H$ becomes of the order of the string scale!
It should be stressed that this result is only  valid when the  the stress tensor is  of the perfect fluid type, {\it i.e.} when it becomes  possible to neglect the  viscosity terms due to mutual and self-interactions. These terms are not negligible when the approximation of small transverse fluctuations breaks down. 

The role played by the transverse fluctuations may be even more dramatic, though. Remember that in the considerations made so far we have supposed that the directions transverse  to our three  brane were static. Assume now that 
these directions are suffering a period of decelerating contraction, while the spatial dimensions of the brane are accelerating. This is -- for instance -- what happens during the stage of radion-driven inflation previously described. 
From Eq. (\ref{exp}) the fluctuations $y_a$ transverse to the brane  
get a  tachyonic mass $
m_y^2= 1/3\left(\ddot{b}/b+3\dot{b}/b\dot{a}/a\right)$,
where now the dots stand for derivatives with respect to the cosmic time $t$.
The situation is very similar to the thermal tachyon which occurs at
the Hagedorn temperature in string theory \cite{sathi,kog}. In that case a condensate of
winding states forms when the periodic imaginary times becomes too
small. In the present case the tachyon forms because of the nontrivial
geometry of the spacetime.  As a given
physical wavelength $ak^{-1}$ of a transverse fluctuation $y_a$  crosses the horizon, the comoving amplitude $h^a_k$  does {\it not}
freeze out; instead  the transverse
physical amplitude
grows {\it faster} than the scale factor $a$ and therefore faster than
the physical coordinates defining the brane. 
 In fact the brane tends to loose
its own identity as the thickness of the brane  grow  faster than the physical coordinates
defining the three brane itself;   the  brane violently fluctuate along
the transverse directions. This phenomenon holds for branes of any dimensionality. If one goes beyond the approximation
of small transverse fluctuations and computes 
the higher order terms in the tachyon effective potential, one can show that -- along with these huge fluctuations --  branes
become almost tensionless \cite{sing1}. This is hardly surprising since the 
brane can   fluctuate along
the transverse directions paying no price in energy. When the  tension  becomes negative, branes become unstable and -- at least in the regime of strong coupling -- it becomes easier             
and easier for the gravitational background to produce further
massless branes. On the other hand, when the transverse sizes
of the branes become larger than the horizon, the branes
presumably break down and decay, but   branes are continuously
created. The rate of brane creation due to these quantum effects 
may be so fast to balance the dilution of the brane density due to the expansion. One might be therefore led into a phase of {\it constant} brane density and exponentially expanding Universe, phase of {\it brane-driven} inflation \cite{inprep} reminescent of the old idea of string-driven inflation \cite{turok}.

The huge transverse fluctuations  of branes in curved spacetime  may be also relevant to the issue
of baryon asymmetry production on our three brane if we accept the idea that there might be  other branes separated from ours
in the extra directions. Even though on our brane global charges, such as the baryon and the lepton number, are conserved with a great accuracy, this is not necessarily the case  on other branes. 
 This breaking might be communicated to our brane by  messenger fields  that leave in the bulk  \cite{sup}. However, this communication
is suppressed by the enormous distance separating the branes in the bulk and this is the reason why today the baryon number  is conserved in 
SM interactions. Nevertheless, at early epochs in the evolution of the Universe the thickness of a brane is highly fluctuating. This makes it possible for two branes separated in the bulk  to overlap.
One can also envisage various other
dynamical phenomena like the melting of two or more branes \cite{sing1}. If our brane was considerably overlapping with another brane where the breaking on the
baryon number is ${\cal O}(1)$, a significant amount of baryon number might have been deposited on our three brane, thus explaining the observed baryon asymmery of our Universe \cite{see}.

The considerations reported here are very preliminary and certainly leave many questions unanswered. For instance, what are the initial conditions for the Universe (or perhaps just for our region of the Universe) right  before the onset of  inflationary PBB stage driven by the radion field? What is the subsequent dynamics of the 
gas of pseudo-Goldstone bosons which is generated  at $H\sim M_s$ and whose presence may help avoiding the singularity?  How does the system get into the familiar radiation-dominated stage? At this point open strings attached to the branes 
have to play a role since  the expansion time $H^{-1}$ is of the order of the typical scale of the stringy network $M_s^{-1}$. It seems safe therefore to assume that this gas of strings attached to the branes and pseudo-Goldstone bosons reaches thermal equilibrium. The fact that fundamental strings have a limiting temperature $\sim M_s$ may be turn out to be crucial  since  it implies  that -- if radiation is created and is in thermal contact with the strings --it cannot attain a density higher than 
$\sim M_s^4$, and the Universe necessarily  is dominated by strings.

We are grateful to T. Banks, R. Brustein, E. Copeland, K. Dienes, M. Dine, S. Dimopoulos, T. Gherghetta, G.F. Giudice, D. Lyth, R. Rattazzi, G. Veneziano and especially to M. Maggiore for
many useful discussions.

\end{document}